\documentclass[12pt]{article}
\usepackage{graphicx}

\def\ignore#1{{}}

\let\oldtheequation=\theequation
\def\doteqs#1{\setcounter{equation}{0}            
\def\theequation{{#1}.\oldtheequation}}
\newcounter{sxn}
\def\sx#1{\addtocounter{sxn}{1} \vskip 1.cm  \goodbreak
\noindent{\large\bf\leftline{\thesxn.~~#1}} \nobreak \vskip -.5cm}
\def\sxn#1{\sx{#1} \doteqs{\thesxn}}

\newcounter{axn}

\date{}

\newdimen\mybaselineskip
\newcommand{\beeq}{\begin{equation}}
\newcommand{\eneq}{\end{equation}}
\newcommand{\beqn}{\begin{eqnarray}}
\newcommand{\eeqn}{\end{eqnarray}}

\def\la{\raise.16ex\hbox{$\langle$}\lower.16ex\hbox{}  }
\def\ra{\, \raise.16ex\hbox{$\rangle$}\lower.16ex\hbox{} }

\def\psibar{ \psi \kern-.65em\raise.6em\hbox{$-$} \lower.6em\hbox{} }
\def\psibarb{ \psi \kern-.65em\raise.6em\hbox{$-$}  }

\begin{document}

\thispagestyle{empty}

\baselineskip=12pt



\vspace*{1.cm}

\begin{center}  
{\LARGE \bf   Validity of the WKB Approximation in Calculating the Asymptotic Quasinormal Modes of Black Holes}
\end{center}

\baselineskip=14pt

\vspace{3cm}
\begin{center}
{\bf  Ramin G.~Daghigh $\sharp$ and Michael D.~Green $\dagger$}
\end{center}

\centerline{\small \it $\sharp$ Natural Sciences Department, Metropolitan State University, Saint Paul, Minnesota, USA 55106}
\vskip 0 cm
\centerline{} 

\centerline{\small \it $\dagger$ Mathematics Department, Metropolitan State University, Saint Paul, Minnesota, USA 55106}
\vskip 1 cm
\centerline{} 

\vspace{1cm}
\begin{abstract}
In this paper, we categorize non-rotating black hole spacetimes based on their pole structure and in each of these categories we determine whether the WKB approximation is a valid approximation for calculating the asymptotic quasinormal modes. We show that Schwarzschild black holes with the Gauss-Bonnet correction belong to the category in which the WKB approximation is invalid for calculating these modes.   In this context, we further discuss and clarify some of the ambiguity in the literature surrounding the validity conditions provided for the WKB approximation.  
\baselineskip=20pt plus 1pt minus 1pt
\end{abstract}

\newpage

\sxn{Introduction}
\vskip 1cm

The equations governing various classes of non-rotating black hole quasinormal mode (QNM) perturbations in radial coordinate $r$ have the general form 
\beeq
\frac{d^2\Psi}{dr^2}+R(r)\Psi=0 ~,
\label{Schrodinger-R}
\eneq 
with the function $R(r)$ of the general form
\beeq
R(r) = {1 \over f(r)^2}\left[\omega^2 -V(r)\right]~,
\label{generic-R}
\eneq
where $\omega$ is the complex QNM frequency with real ($\omega_R$) and imaginary ($\omega_I$) parts, $V(r)$ is a function that depends on the type of perturbation and the spacetime structure, and $f(r)$ is the black hole metric function that appears in the spcetime metric
\beeq
ds^2=-f(r)dt^2+{dr^2 \over f(r)}+r^2d\Omega^2_{D-2}~
\label{GB-metric}
\eneq
with the spacetime dimension $D$.

Methods developed to calculate the highly damped\cite{Motl2, Andersson, QNM1}  and other asymptotic\cite{Natario, highly-real} QNM frequencies of black holes (see also \cite{QNM3} for a comprehensive review of black hole QNM's) are based on the WKB approximation.  Here, by highly damped we mean $|\omega| \approx |\omega_I| >> \omega_R $ and by asymptotic we mean when $|\omega| \rightarrow \infty $ in general.  The highly damped QNM's have attracted considerable attention in the literature as a result of the work of Hod\cite{Hod} and more recently Maggiore\cite{Maggiore}, who have attempted to link these modes to the horizon area quantization of black holes.  Also, in a recent paper\cite{PK-Ramin}, a link between the highly damped QNM's and the small scale structure of black hole spacetimes has been established.  The authors of \cite{PK-Ramin} show that the highly damped QNM's are sensitive to any small scale (for example a small charge or angular momentum) that is added to a Schwarzschild metric.  This includes any additional length scale due to a quantum correction. The authors of \cite{PK-Ramin} demonstrate this explicitly by calculating the highly damped QNM's of the quantum corrected black hole derived in \cite{pk09}.  The results of this paper are also relevant to the analytic calculations of the greybody factors for black holes at large imaginary frequencies.  See, for example, \cite{greybody}.

The two approximate, linearly independent, complex, WKB solutions to the differential equation (\ref{Schrodinger-R}) are
\beeq
\left\{ \begin{array}{ll}
                   \Psi_1^{(t)}(r)=Q^{-1/2}(r)\exp \left[+i\int_{t}^rQ(r')dr'\right]~\\
                   \\
                   \Psi_2^{(t)}(r)=Q^{-1/2}(r)\exp \left[-i\int_{t}^rQ(r')dr'\right]~,
                   \end{array}
           \right.        
\label{WKB}
\eneq
where $Q^2=R$ and $t$ is customarily taken to be a zero of $Q$ in the complex $r$-plane.  
In the methods developed in \cite{Motl2, Andersson, QNM1, Natario, highly-real}, one needs to analytically continue the complex WKB solutions (\ref{WKB}) into the whole complex coordinate plane.  After locating the poles and zeros of the function $R(r)$, and consequently $Q(r)$, one is able to determine the topology of certain lines called Stokes and anti-Stokes lines. Along Stokes lines the phase of the WKB solutions ($\int Q dr$) is purely imaginary and along anti-Stokes lines this phase is purely real.  In order to determine a WKB condition on the QNM frequency $\omega$, in most cases one needs to find a closed contour around the pole at the event horizon.  We choose this contour along anti-Stokes lines because the solutions do not change character on these lines.  In other words, if we know the behavior of the solution on a point of an anti-Stokes line, then we know the behavior of the solution everywhere on that line.  Along this contour, one always encounters some of the poles of the function $R(r)$ in the complex plane.\footnote{In the method developed in \cite{Motl2}, one does not need to determine the topology of Stokes lines or the position of the zeros of $Q$.  For more details see the end of Sec.\ 4.}   The details of these calculations are not important for the purpose of this paper.  What is important is that, in the infinite damping limit, it is necessary to approach infinitely close to some of these poles.  This can be realized by noticing that in the limit $|\omega | \rightarrow \infty$, the function $V(r)$ can be neglected everywhere in the complex plane compared to $\omega^2$ except when we are infinitely close to the poles of $V(r)$.   Consequently, one needs to make sure that the WKB approximation is valid in the region near those poles.  

The rest of this paper proceeds as follows.  In Sec.\ 2, we provide a generic function $R(r)$ which includes the pole structure of all types of non-rotating black hole spacetimes (at least the ones we are aware of).  In Sec.\ 3, we categorize black hole spacetimes into four categories and in each of these categories we determine whether the WKB approximation is valid for calculating the asymptotic QNM frequencies.  In Sec.\ 4, we show that the WKB approximation is not valid for calculating the highly damped QNM's of  Schwarzschild black holes with the Gauss-Bonnet correction.  In Sec.\ 5, we discuss additional WKB validity conditions in the context of the black hole QNM problem.

\sxn{A Generic Example}
\vskip 1cm

Let us assume that $R$ has the generic form 
\beeq
R(r) = F(r)-\frac{A}{(r-r_*)^a}~,
\label{general-R}
\eneq
where $r_*$ is the location of one of the poles of the function $R(r)$ in the complex plane, $A$ is a real or complex constant, and $a$ is a real number greater than zero.  We assume that the function $F(r)$ is a general function that can have a pole at the same location of $r=r_*$ but with a power less than $a$ in its denominator.  We want to check if the WKB approximation is valid in the region where $r\rightarrow r_*$.   A well known condition for checking the validity of the WKB approximation is:
\beeq
|R^{-\frac{3}{2}}\frac{dR}{dr}|<<1~.
\label{WKBcondition}
\eneq
There exist some ambiguity in the literature regarding the extent of the applicability of the above condition.  
Many books \cite{Bransden, Kirsten, Liboff, white} interpret (\ref{WKBcondition}) as a condition that the function $R$ be ``slowly varying'' and {\em suggest} it is sufficient to guarantee accuracy of the WKB approximation. 
However, in \cite{Northover} this condition is only one of two conditions required for sufficiency.
Perhaps more accurately, condition (\ref{WKBcondition}) is referred to as a ``rule of thumb'' in \cite{Manoukian}.  
In this section, we will discuss the extent of the validity of this condition in the context of the generic example provided in Eq.\ (\ref{general-R}).  We will discuss other WKB validity conditions in Sec.\ 5.

For $a>2$, it is easy to show that near $r=r_*$ the validity condition in (\ref{WKBcondition}) is satisfied. In order to further make sure that the WKB approximation is valid for $a>2$, we need to check if, in the asymptotic region $r \rightarrow r_*$, the solution to the wave equation (\ref{Schrodinger-R}) matches the WKB solutions (\ref{WKB}).  Therefore, we solve the wave equation (\ref{Schrodinger-R}) with the function $R(r)$ given in (\ref{general-R}) in the limit $r\rightarrow r_*$.  Note that, in this limit, $F(r)$ can be neglected.  The solution is 
\beeq
\Psi(r) \approx  C_1 \sqrt{r-r_*}J_{\frac{1}{2-a}}\left[\frac{2}{2-a}\sqrt{-A}(r-r_*)^\frac{2-a}{2}\right]+C_2 \sqrt{r-r_*}J_{-\frac{1}{2-a}}\left[\frac{2}{2-a}\sqrt{-A}(r-r_*)^\frac{2-a}{2}\right] ~,
\label{exact-solution}
\eneq
where $C_{1,2}$ are constants and $J_\nu$ is the Bessel function of the first kind.  On the other hand, the WKB solutions (\ref{WKB}) behave like
\beeq
\Psi_{WKB}(r)\approx  C_+ (r-r_*)^{\frac{a}{4}}e^{i \frac{2}{2-a}\sqrt{-A}(r-r_*)^\frac{2-a}{2}}+C_- (r-r_*)^{\frac{a}{4}}e^{-i\frac{2}{2-a}\sqrt{-A}(r-r_*)^\frac{2-a}{2}}~          
\label{WKB-a}
\eneq
when $r\approx r_*$.
For the WKB approximation to be valid near $r=r_*$, the solutions (\ref{exact-solution}) and (\ref{WKB-a}) have to match.  To check this, we use the asymptotic behavior of the Bessel function $J_\nu (x)$ when $x>>1$, i.e. 
\beeq
J_\nu (x)\approx \sqrt{2 \over \pi x}\cos\left(x-{\nu \pi \over 2}-{\pi \over 4}\right)~.
\label{asymptotic-J}
\eneq
Using the above asymptotic behavior, we can easily show that solution (\ref{exact-solution}) can be reduced to the WKB solution (\ref{WKB-a}).  Here, we have to point out that the asymptotic behavior (\ref{asymptotic-J}) can only be used when the argument of the Bessel function has a real value.  Note that the argument of the Bessel function in (\ref{exact-solution}) is equal to the phase ($\int Q dr$) of the WKB solutions.  Since at every point in the complex $r$-plane we can find a curve on which  $\int Q dr$ is purely real, we can always use the asymptotic behavior (\ref{asymptotic-J}).

Things change for the case of $a=2$.  In this case, when $r\rightarrow r_*$, we have 
\beeq
\left |R^{-\frac{3}{2}}\frac{dR}{dr}\right |=\left |{2\over \sqrt{A}}\right |~,
\eneq
which tells us that the condition (\ref{WKBcondition}) is only satisfied if $|A|>>1$.  However, we can still use the WKB approximation if we apply a modification.  It is easy to show that, when $r\approx r_*$, the solutions of the wave equation (\ref{Schrodinger-R}) with $R$ given by (\ref{general-R}) are 
\beeq
\Psi(r)\propto (r-r_*)^{\frac{1}{2}\pm {\sqrt{1+4A}\over 2}}~.
\eneq
Meanwhile, if we take $Q^2=R$, we get
\beeq
Q^{-\frac{1}{2}}\propto (r-r_*)^\frac{1}{2}~
\eneq
and 
\beeq
\int Q dr \approx \pm i \sqrt{A} \ln (r-r_*)~,
\eneq
which means the WKB solutions are
\beeq
\Psi_{1,2} \propto (r-r_*)^{\frac{1}{2}\pm \sqrt{A}}~.
\eneq
Therefore, the WKB solutions do not have the correct behavior near $r=r_*$.  This discrepancy, however, can be resolved by choosing 
\beeq
Q^2=R-\frac{1}{4(r-r_*)^2}~.
\label{shift}
\eneq
This modification is very similar to the Langer modification used in the WKB approximation of radial quantum problems.
It is important to point out that the above shift does not change the topology of anti-Stokes lines, which are crucial in the asymptotic QNM frequency calculations.  

In the case of $a<2$, condition (\ref{WKBcondition}) is not satisfied near $r=r_*$.   The fact that the WKB approximation is not valid for $a<2$ can be observed by noticing that the solution (\ref{exact-solution}) approaches
\beeq
\Psi(r) \approx    \bar{C}_1 (r-r_*) + \bar{C}_2 ~
\eneq
in the $r\rightarrow r_*$ limit.  But, in the same limit, the WKB solution (\ref{WKB-a}) approaches
\begin{eqnarray}
\Psi_{WKB}(r) \approx  \bar{C}_+  (r-r_*)^{\frac{a}{4}} + \bar{C}_-  (r-r_*)^{\frac{a}{4}}~.               
\end{eqnarray}
It is clear that the WKB solutions do not have the correct behavior near these poles and their behavior cannot be corrected by a simple shift of the form (\ref{shift})
in which the topology of anti-Stokes lines is unaltered.  To be more precise, in order to match the solutions of the QNM wave equation and the WKB solutions near $r=r_*$, we need to take 
\beeq
Q^2=\frac{R}{(r-r_*)^{\frac{5}{2}}}~
\eneq 
if the constant $\bar{C}_2$ is zero.  Otherwise, we need the shift
\beeq
Q^2=R{(r-r_*)^{\frac{3}{2}}}~.
\eneq 
Such shifts will modify the topology of Stokes/anti-Stokes lines since the pole structure is altered.

This example shows that if one accounts for the Langer-type modification 
given in (\ref{shift}), then the WKB validity condition (\ref{WKBcondition}) is a sufficient condition.  However, if this validity condition is not satisfied, it may still be possible to use the WKB approximation.

\sxn{Categorizing Black Hole Spacetimes for QNM Calculations}
\vskip 1cm

Based on the generic function $R(r)$ provided in Eq.\ (\ref{general-R}), we can categorize non-rotating black holes based on their pole structure into four different categories: 
 
1) At all poles $r_*$, $a > 2$. 

2) At all poles $r_*$, $a =2$. 

3) At all poles $r_*$, $a < 2$. 

4) Some of the poles have $a \ge 2$ and others have $a < 2$.

\noindent Some black hole examples of the second category are Schwarzschild, Reissner-Nordstr$\ddot{\rm{o}}$m, Schwarzschild-de Sitter, Schwarzschild-anti de Sitter and the non-singular quantum corrected black hole spacetime derived in \cite{pk09}.  We are not aware of any example of the first or the third category.  The Schwarzschild black hole with the Gauss-Bonnet correction, which will be explained in the next section, is an example of the fourth category.  The WKB approximation in calculating the asymptotic QNM frequency is always valid for the first and second category and it is always invalid for the third category 
.  The validity of the WKB approximation in the fourth category depends on whether we encounter a pole  with $a < 2$ along the anti-Stokes lines encircling the event horizon or not.  If we do, then the WKB approximation is invalid.  Otherwise, it is still fine to use this approximation. 

In this paper we have only talked about non-rotating black holes.  For rotating black holes, similar categorization may be possible. For example, the pole structure of Kerr black holes is of category 2, but in the case of Kerr black holes the asymptotic behavior of the wave equation near the poles is irrelevant in the high damping limit because the contour that we take along anti-Stokes lines never encounters any pole in the complex plane\cite{Keshet-Hod}.  Therefore, in the Kerr case, we do not need to check if the WKB approximation is valid near the poles.  We predict that the same should be true for other types of rotating black holes such as Kerr-de Sitter, Kerr-anti de Sitter, etc.  

\sxn{The Case of Gauss-Bonnet}
\vskip 1cm
Schwarzschild black holes with Gauss-Bonnet (G-B) corrections\cite{GB1, GB2, GB3} (hereafter G-B black holes) are black holes in spacetime dimensions greater than four which appear when one includes the leading order higher curvature terms arising in the low energy limit of string theories \cite{Zwiebach}.   

The authors of \cite{GB-Daghigh} use a combination of analytic and numeric techniques based on the WKB approximation to calculate the QNM frequencies of G-B black holes in the intermediate damping range of  
\beeq
1<<|\omega|\mu^{\frac{1}{D-3}}<< \left[\frac{4(D-2)^2-1}{(8\alpha \mu)^\frac{2D-4}{D-1}}\right]^\frac{1}{2} \mu^{\frac{D-2}{D-3}}~,
\label{inter-damping}
\eneq  
where $D$ is the spacetime dimension, $\alpha$ is the G-B coupling constant and the parameter $\mu$ is related to the ADM mass $M$ given by
\beeq
M = {(D-2)A_{D-2} \over 8 \pi G_D}\mu~.
\eneq
Working in the intermediate damping region mentioned above allows the authors of \cite{GB-Daghigh} to do the calculations in a region of the complex $r$-plane where $r>>(-8\alpha \mu)^\frac{1}{D-1}$.  In this region, the calculations are not affected by the fine structure of the extra poles in the complex $r$-plane, which appear as a result of the G-B correction.  These poles are located at the roots of $r^{D-1}+8\alpha \mu=0$ and $(D-3)r^{D-1}+4\alpha \mu(D-5)=0$.   Under such circumstances it was shown in \cite{GB-Daghigh} that, in the intermediate damping region (\ref{inter-damping}),
\beeq
2\pi \frac{2\mu+\alpha}{\sqrt{2\mu-\alpha}}\omega_R \approx \ln (2.99926)~
\eneq
when $D=5$. 
This result is very close to the Schwarzschild result of $\ln(3)$\cite{Motl2}. In the limit of $\alpha \rightarrow 0$, the correct Schwarzschild value of $\ln(3)$ is recovered.  For further discussions on the Schwarzschild limit of the highly damped QNM's of different black holes, see \cite{Andersson} and \cite{PK-Ramin}.  In the highly damped limit, where 
\beeq
|\omega|>> \left[\frac{4(D-2)^2-1}{(8\alpha \mu)^\frac{2D-4}{D-1}}\right]^\frac{1}{2} \mu~,
\label{high-damping}
\eneq 
the authors of \cite{GB-Daghigh} found that the standard analytic techniques cannot be applied in a straightforward way.
In what follows we show that, in fact, the WKB approximation is an invalid approximation for calculating the QNM's of G-B black holes in the highly damped region (\ref{high-damping}).

We can use the metric function for a Schwarzschild black hole with the G-B correction\cite{GB1}
\beeq
f(r)=1+{r^2\over 2\alpha}-{r^2\over 2\alpha}\sqrt{1+{8\alpha\mu \over r^{D-1}}}~
\label{GB-f}
\eneq
to find that the function $R(r)$ in five spacetime dimensions can be approximated in the $|\omega| \rightarrow \infty$ limit by\cite{GB-Daghigh} 
\beeq
R(r)\sim \frac{\omega^2}{f^2} -{35 \over 4r^2}+ {48 \alpha \mu \over (r^4+8 \alpha \mu)^{3/2}}~.
\label{Rr5d}
\eneq
This function has a pole structure of category 4, where for one pole $a=2$ and for the other poles $a=3/2$.  Therefore, according to the results obtained in Sec.\ 2, the WKB approximation would be invalid if one encounters a pole with $a=3/2$ along anti-Stokes lines that loop around the event horizon.  This turns out to be the case for the G-B black hole QNM's in the highly damped region (\ref{high-damping}).  The path along anti-Stokes lines for the G-B case in the high damping limit is shown in Figure 1 of reference \cite{GB-Daghigh}.  In the intermediate damping region (\ref{inter-damping}), however, the path that we take along anti-Stokes lines around the event horizon never passes near the poles at $r=(-8\alpha \mu)^{1/4}$.  Therefore, the calculations can be done in the intermediate damping region using the WKB approximation.  The path along anti-Stokes lines for the G-B case in the intermediate damping region can be found in Figure 4 of reference \cite{GB-Daghigh}.

We also would like to point out that in the method developed by Motl and Neitzke\cite{Motl2} for calculating the highly damped QNM's, determining the location of the zeros of $Q$ or the fine structure of Stokes and anti-Stokes lines near the poles is not necessary.  In this method, the wave equation near the poles is solved analytically and then the solutions are matched to the WKB solutions along anti-Stokes lines when one moves away from the poles.  In the region close to the poles where $r^4 \rightarrow -8\alpha \mu$ the wave equation can be approximated as
\beeq
\frac{d^2\Psi}{dr^2}+\left[-\frac{\alpha \omega^2}{2\mu}+ {48 \alpha \mu \over (r^4+8 \alpha \mu)^{3/2}} \right]\Psi=0 ~
\label{GB-wave-eq}
\eneq 
The WKB validity condition in (\ref{WKBcondition}) is not satisfied for the above wave equation because
\beeq
\left |R^{-\frac{3}{2}}\frac{dR}{dr}\right|=\left| \left[-\frac{\alpha \omega^2}{2\mu}+ {48 \alpha \mu \over (r^4+8 \alpha \mu)^{3/2}} \right]^{-3/2} \frac{-288\alpha \mu}{(r^4+8\alpha \mu)^{5/2}} \right| \propto  \left| (r^4+8\alpha \mu)^{-1/4}\right | \gg 1~
\eneq 
near the poles at $r= (8\alpha \mu)^{1/4}$.
Unfortunately, Eq.\ (\ref{GB-wave-eq}) does not have a simple solution and we cannot show directly if the solutions to this differential equation can be matched to the WKB solutions (\ref{WKB}), but the fact that the validity condition (\ref{WKBcondition}) is not satisfied is a strong indication that the WKB approximation is invalid.  In addition, since the two methods in \cite{Motl2} and \cite{Andersson} should produce consistent results, it would be highly unlikely that in one method the WKB approximation is valid and in the other invalid.

\sxn{Other Validity Conditions for the WKB Approximation}
\vskip 1cm

To discuss some of the other validity conditions used for the WKB approximation in the literature, it is useful to review the derivation below.  For more details see \cite{bender}.  

In the WKB theory, we assume a solution to the wave equation (\ref{Schrodinger-R}) in terms of an exponential power series of the form
\beeq
\Psi(r) \sim \exp\left[ \frac{1}{\delta} \sum_{n=0}^\infty \delta^n S_n(r) \right] ~
\label{asymptotic-expansion}
\eneq 
when $\delta \rightarrow 0$.  
The WKB series $\sum \delta^n S_n(r)$ is divergent for most cases and for that reason the asymptotic notation $\sim$ is used in (\ref{asymptotic-expansion}) instead of the equality sign ($=$).
By inserting (\ref{asymptotic-expansion}) into the wave equation (\ref{Schrodinger-R}), we obtain
\beeq
\frac{1}{\delta^2}S_0'^2+\frac{2}{\delta}S_0'S_1'+\frac{1}{\delta}S_0''+...=-R(r) ~.
\label{derivative-terms}
\eneq 
The first term on the left is the dominant term as $\delta \rightarrow 0$ and must have the same order of magnitude as $R(r)$.  Therefore, we get 
\beeq
\frac{1}{\delta^2}S_0'^2=-R(r) ~.
\eneq 
The coefficients of $\delta^i$, $i > -2$, in (\ref{derivative-terms}) have to vanish.  This leads to
\beeq
2S_0'S_1'+S_0''=0 ~,
\eneq 
and
\beeq
2S_0'S_n'+S_{n-1}''+\sum_{j=1}^{n-1} S_j'S_{n-j}'=0~,~~~n\ge 2~.
\eneq 
Solving these equations gives
\beeq
\frac{1}{\delta} S_0(r)=\pm \int ^r \sqrt{-R(t)} dt ~,
\eneq 
\beeq
S_1(r)=-\frac{1}{4} \ln [-R(r)] ~,
\eneq 
\beeq
\delta S_2(r)=\pm \int^r \left[ \frac{R''}{8(-R)^{3/2}}+\frac{5R'^2}{32(-R)^{5/2}}\right] dt ~,
\label{S2}
\eneq 
and so on.  The WKB approximation used in (\ref{WKB}) is just $\exp [\frac{1}{\delta} S_0+S_1]$, which is sometimes called the leading-order WKB approximation.

For the WKB approximation to be valid on an interval, the series $\sum \delta^n S_n(r)$  has to be an asymptotic series in $\delta$ as $\delta \rightarrow 0$ uniformly for all $r$ on the interval.   This requires that 
\beeq
|S_1(r)| << \left | \frac{1}{\delta}S_0(r) \right |~~\mbox{as}~~\delta \rightarrow 0
\label{cond-1}
\eneq 
and
\beeq
|\delta S_2(r)| << |S_1(r)|~~\mbox{as}~~\delta \rightarrow 0,
\label{cond-2}
\eneq 
which have to hold uniformly in $r$.  
We also need to require that the first term we drop is small, i.e.
\beeq
|\delta S_2(r)| << 1~~\mbox{as}~~\delta \rightarrow 0.
\label{cond-3}
\eneq 
This is because the WKB series appear in the exponent and if the above condition is not satisfied the error can be large.  Here we are only concerned with the leading-order WKB approximation, which means that the conditions can stop at $S_2$ and no further conditions are necessary.  

The validity condition (\ref{WKBcondition}), which was mentioned in Sec.\ 2, implies the condition (\ref{cond-3}).  This can be shown by rewriting Eq.\ (\ref{S2}) as
\beeq
\delta S_2(r)=\mp  \frac{R'}{8(-R)^{3/2}} \pm \int^r  \frac{R'^2}{32(-R)^{5/2}}  dt ~.
\eneq 
It is now easy to see that $|\delta S_2| << 1$ as long as condition (\ref{WKBcondition}) is satisfied.  It is also possible to see where (\ref{WKBcondition}) comes from by realizing that the dominant term, $S_0'^2/\delta^2$, in Eq.\ (\ref{derivative-terms}) has to be much larger than either one of the terms of $O\left(\frac{1}{\delta}\right)$ as $\delta \rightarrow 0$.

It is interesting to point out that, similar to the validity condition (\ref{WKBcondition}), the sets of validity conditions (\ref{cond-1}) through (\ref{cond-3}) are all satisfied for category 1 black holes and fail for category 3.  For category 2 black holes, they may or may not be satisfied depending on how large or small the constant $A$ is in Eq.\ (\ref{generic-R}).  In such a situation, a Langer-type modification will allow us to use the WKB approximation.




\newpage 


\def\jnl#1#2#3#4{{#1}{\bf #2} #3 (#4)}

\def\Zphys{{\em Z.\ Phys.} }
\def\jssc{{\em J.\ Solid State Chem.\ }}
\def\jpsJ{{\em J.\ Phys.\ Soc.\ Japan }}
\def\ptps{{\em Prog.\ Theoret.\ Phys.\ Suppl.\ }}
\def\PTP{{\em Prog.\ Theoret.\ Phys.\  }}
\def\LNC{{\em Lett.\ Nuovo.\ Cim.\  }}

\def\JMP{{\em J. Math.\ Phys.} }
\def\NPB{{\em Nucl.\ Phys.} B}
\def\NP{{\em Nucl.\ Phys.} }
\def\PLB{{\em Phys.\ Lett.} B}
\def\PL{{\em Phys.\ Lett.} }
\def\PRL{\em Phys.\ Rev.\ Lett. }
\def\PRB{{\em Phys.\ Rev.} B}
\def\PRD{{\em Phys.\ Rev.} D}
\def\PR{{\em Phys.\ Rev.} }
\def\PRe{{\em Phys.\ Rep.} }
\def\AP{{\em Ann.\ Phys.\ (N.Y.)} }
\def\RMP{{\em Rev.\ Mod.\ Phys.} }
\def\ZPC{{\em Z.\ Phys.} C}
\def\SCI{\em Science}
\def\CMP{\em Comm.\ Math.\ Phys. }
\def\MPLA{{\em Mod.\ Phys.\ Lett.} A}
\def\IJMPA{{\em Int.\ J.\ Mod.\ Phys.} A}
\def\IJMPB{{\em Int.\ J.\ Mod.\ Phys.} B}
\def\cmp{{\em Com.\ Math.\ Phys.}}
\def\JPA{{\em J.\  Phys.} A}
\def\CQG{\em Class.\ Quant.\ Grav.~}
\def\ATMP{\em Adv.\ Theoret.\ Math.\ Phys.~}
\def\PRSA{{\em Proc.\ Roy.\ Soc.} A }
\def\ibid{{\em ibid.} }
\vskip 1cm

\leftline{\bf References}

\renewenvironment{thebibliography}[1]
        {\begin{list}{[$\,$\arabic{enumi}$\,$]}  
        {\usecounter{enumi}\setlength{\parsep}{0pt}
         \setlength{\itemsep}{0pt}  \renewcommand{\baselinestretch}{1.2}
         \settowidth
        {\labelwidth}{#1 ~ ~}\sloppy}}{\end{list}}



\begin{thebibliography}{99}
\small
\baselineskip=16pt



\bibitem{Motl2}
L.\ Motl, \jnl{\ATMP}{6}{1135}{2003}; L.\ Motl and A.\ Neitzke, \jnl{\ATMP}{7}{307}{2003}.

\bibitem{Andersson}
N.\ Andersson and C.\ J.\ Howls, \jnl{\CQG}{21}{1623}{2004}.


\bibitem{QNM1}
A.\ Maassen van den Brink, \jnl{\JMP}{45}{327}{2004}; S.\ Musiri and G.\ Siopsis, \jnl{\CQG}{20}{L285}{2003}.


\bibitem{Natario} 
V.~Cardoso, J.~Natario and R.~Schiappa, \jnl{\JMP}{45}{4698}{2004}; J.~Natario and R.~Schiappa, \jnl{\ATMP}{8}{1001}{2004}.

\bibitem{highly-real} 
R.\ G.\ Daghigh and M.\ D.\ Green,  \jnl{\CQG}{26}{125017}{2009}; R.\ G.\ Daghigh, JHEP 0904:045 (2009). 
 


\bibitem{QNM3}
E.\ Berti,  V.\ Cardoso and A.\ O.\ Starinets, \jnl{\CQG}{26}{163001}{2009}.


\bibitem{Hod}
S.\ Hod, \jnl{\PRL}{81}{4293}{1998}.

\bibitem{Maggiore} M.\ Maggiore, \jnl{\PRL}{100}{141301}{2008}. 

\bibitem{PK-Ramin}
J.~Babb, R.~G.~Daghigh, G.~Kunstatter, \jnl{\PRD}{84}{084031}{2011}. 

\bibitem{pk09} A.~Peltola and G.~Kunstatter, \jnl{\PRD}{79}{061501 (R)}{2009}; \jnl{\PRD}{80}{044031}{2009}. 

\bibitem{greybody}
T.\ Harmark,  J.\ Natario,  R.\ Schiappa,  \jnl{\ATMP}{14}{727}{2010}; arXiv:0708.0017 [hep-th].


\bibitem{Bransden} B.\ H.\ Bransden and C.\ J.\  Joachain, {\em Quantum Mechanics}, 2nd ed., Prentice Hall, 2000.

\bibitem{Kirsten} H.\ J.\ W.\ M\"uller-Kirsten, {\em Introduction to Quantum Mechanics, Schr\"odinger Equation and Path Integral}

\bibitem{Liboff} R.\ L.\ Liboff, {\em Introductory Quantum Mechanics}, 4th ed., Addison Wesley, 2003.

\bibitem{white} R.\ White, {\em Asymptotic Analysis of Differential Equations}, Imperial College Press, 2005.


\bibitem{Northover} F.\ H.\ Northover, \jnl{\JMP}{10}{715}{1969}

\bibitem{Manoukian} E.~B.~Manoukian, {\em Quantum Theory A Wide Spectrum}, Springer, 2006.




\bibitem{Keshet-Hod}U.\ Keshet and S.\ Hod, \jnl{\PRD}{76}{061501}{2007};  U.\ Keshet and A.\ Neitzke, \jnl{\PRD}{78}{044006}{2008}. 


\bibitem{GB1}D.~G.~Boulware and S.\ Deser, \jnl{\PRL}{55}{2656}{1985}.
\bibitem{GB2}J.~Wheeler, \jnl{\NPB}{268}{737}{1986}.
\bibitem{GB3}D.~L.~Wiltshire, \jnl{\PRD}{38}{2445}{1988}.

\bibitem{Zwiebach}B.~Zwiebach, \jnl{\PLB}{156}{315}{1985}.

\bibitem{GB-Daghigh}
R.~G.~Daghigh, G.~Kunstatter and J.~Ziprick, \jnl{\CQG}{24}{1981}{2007}.


 







\bibitem{bender} C.\ Bender and S.\ Orszag, {\em Advanced Mathematical Methods For Scientists And Engineers}, McGraw-Hill, 1978.


\end{thebibliography}
\end{document}